# High Temperature Superconductivity In Pd-[H(D)]$_X$ System

## K.P. Sinha

*Department of Physics, Indian Institute of Science, Bangalore 560 012*
*D-105, Sterling Residence, Dollar colony, Bangalore 560 094*
*e.mail: kpsinha@gmail.com*

**Abstract**

A theoretical model involving phonon and lochon (local charged boson) mediated pairing of electrons in Pd(H/D)$_x$ is suggested to explain the possible observation of superconductivity in the range 51K$\leq$T$_c$$\leq$295K (when x exceeds unity). In the combined mechanism the lochons have transient existence on (H$^+$/D$^+$) resulting in (H$^-$/D$^-$) species. The coupling parameter $\lambda_l$ for lochon induced processes depend linearly on x. The model explains the observation of the enhancement of T$_c$ for x varying from 0.6 to 1.6. While the experimental observations are of great importance, it is desirable that other experimental groups confirm the results.

**Introduction**

Palladium hydride has been known to be a superconductor from 1972 but in the low temperature regime with the critical $T_c$ ranging from 6K to 9K for the ratio x = H/Pd varying from 0.75 to 1[1,2]. Palladium is loaded with hydrogen (H) or its isotope deuterium (D) by diffusion from gas or by electrolytic means. H or D enter interstitial sites in the host (Pd) (f.c.c.) lattice leading to an expansion of lattice structure[3]. Increased loading by H(D) produces displacement of host metal atoms resulting in vacant host sites. This renders the deformed host lattice very mobile for H(D) which shed their electrons to the 4d shell of Pd which has 0.36 positive holes with the same number of electrons in the 5 sp band[4].

If one takes into account that in the f.c.c. lattice of Pd there are twelve interstitial sites (taking both octahedral and tetrahedral) which can be occupied by H(D), the value of x can be boosted to 3. This ideal situation motivated some workers in this field to push H(D) loading of Pd beyond x = 1[5]. They expected to achieve still higher superconducting critical temperature $T_c$ in Pd-H(D)$_x$ system with high and stable x. Tripodi et al. claim to have produced PdH$_x$ system with x values upto 1.7 by using an electrochemical set up followed by a stabilization process which maintains a stable hydrogen concentration in Pd lattice[5-7]. Their measurement of the normalized resistance (R/R$_o$) (where Ro is the resistance of pure Pd) shows the superconducting critical temperature $T_c$ for various values of x range from 51.6 K to 272.5 K. They predicted that a critical temperature $T_c$ = 300 K is achievable for $x_c$ = 1.6. From the experimental data of low $T_c$ systems and using their high $T_c$ data a relation is extrapolated which gives $T_c = nx_n$ with n = 7.86[6].

The fact that pure Pd is not a superconductor and superconductivity appears in Pd(H,D)$_x$ system only when x ≥ 0.6 suggests an important role of H(D) atoms. The role of optical phonons involving hydrogen (deuterium) atoms or ions has been invoked for low $T_c$ systems. However, phonon mechanism alone cannot account for $T_c$ higher than 30K[8].

In the context of Pd(H-D)$_x$ system it is to be noted that the electron of the H or D goes into the d shell of the system. Thus the majority of H(D) will be in the ionized state H$^+$ or D$^+$.

In what follows, a model involving double charge fluctuation leading to transient formation of local charge bosons on H$^-$(D$^-$) is developed which gives an important electronic mechanism (in addition to phonon mechanism) of superconductivity in Pd[H)D)]$_x$ systems.

**Theoretical Concepts**



The stabilization of real space singlet pair of electrons via transient transition to empty orbital has been invoked a long time back[9]. This mechanism has been successfully extended to superconductivity in electron as well as hole type systems in the past decades[10-14]. The formation of pre-formed pairs due to this mechanism has found applications in high $T_c$ superconductors, namely cuprates[15-17].

The presence of such centres which can harbour such pairs can lead to the pairing of fermions (whether electrons or holes) via the process.

$(\underline{k}\uparrow)-X^o(-\underline{k}\downarrow) \rightarrow (\ ) - X^{\uparrow\downarrow} - (\ ) \rightarrow (\underline{k}^1\uparrow) - X^o - (-\underline{k}^1\downarrow)$     (A)

(The states $X^o$ and $X^{\uparrow\downarrow}$ essentially represent a two-level system).

The above channel is in addition to the phonon mediated interaction channel. The combined (phononic and electronic) mechanism can lead to high $T_c$ systems.

For Pd(H/D)$_x$ system the H/D exist in ionized state $H^+/D^+$. However, strong electron-optical phonon interaction involving $H^+/D^+$ ions reduces the Coulomb repulsion between two electrons in the 1S atomic orbital. Thus, the appearance of transient state $H^-/D^-$ becomes a real possibility. Energetically, $H^-/D^-$ will thus be more stable than H/D. The electron pair on $H^-/D^-$ form a small on site localized pair, a kind of composite boson. These local charged bosons (lochons) can give rise to the $\underline{k}$-space pairing of conduction electrons. In the following sections, we give a formal picture of this combined mechanism involving phonons and local pairs.

**Theoretical Formulation And Results**

The Hamiltonian comprising the conduction electrons of the system Pd(H/D)$_x$ and the two level systems of the empty ($H^+/D^+$) and doubly occupied ($H^-/D^-$) and their interactions can be written as

$H = \sum E_{\underline{k}} C^+_{\underline{k}\sigma} C_{\underline{k}\sigma} + \sum h\Omega_l b^+_l b_l$

$\quad + \sum G_l (C^+_{\underline{k},1\sigma1} C^+_{\underline{k}2\sigma2} b_l + C_{k2\sigma2} C_{k,1\sigma1} b^+_l)$

$\quad + H_{ph} + H_{ep} + H_c,$         (1)

where $E_{\underline{k}}$ is the single particle energy of electrons in the conduction band, $C^+_{\underline{k}\sigma}$, $C_{\underline{k}\sigma}$ the usual creation, annilihation operators in the state $|\phi_{\underline{k}\sigma}\rangle$, $\underline{k}$ being the wave vector and $\sigma$ the spin index. The second term is the energy operator for the two level Boson system and $b^+_l$, $b_l$ the corresponding composite operator defined as

$b^+_l = C^+_{l\uparrow} C^+_{l\downarrow}$     , $b_l = C_{l\downarrow} C_{l\uparrow}$         (2)



which create ($b^+_l$) and annihilate ($b_l$) electron pairs at site l in state $|\phi_l\rangle$ and in the spin singlet state.

$$h\Omega_l = E_l(X_l^{\uparrow\downarrow}, Q_l) - E_l(X^o_l, Q^o_l), \qquad (3)$$

being the energy difference between the two levels of the charged boson systems[15,16].

$$G_l = \langle \phi_{\underline{k}1}\phi_{\underline{k}2} | V(12) | \phi_l\phi_l \rangle \qquad (4)$$

is the two body correlated transfer matrix element involving doubly occupied orbial $|\phi_l\rangle$ and single occupied states $|\phi_{\underline{k}\sigma}\rangle$ of the effective two particle operator $V(12)$. The form of this interaction is different from the usual electron-phonon interaction which involves scattering of an electron from state $|\phi_{\underline{k}\sigma}\rangle$ to $|\phi_{\underline{k}!\sigma}\rangle$. In the relation (4), we have the destruction of a pair of electrons followed by the creation of local pairs (a composite of two fermions i.e., a charged boson). In (1) $H_{ph}$ is the usual phonon and $H_{ep}$ the electron phonon part of the Hamiltonian[18]; the explicit forms of these are not written here. The effects will be incorporated in the final expression along with the screened Coulomb repulsion between two conduction electrons.

The final form of the pairing interaction is obtained by methods discussed earlier[9-18]. We have

$$H = \sum E_{\underline{k}} C^+_{\underline{k}\sigma} C_{\underline{k}\sigma} - \sum (V_{ph} + V_l(x) - V_{sc}) C^+_{\underline{k}\uparrow} C^+_{-\underline{k}\downarrow} C_{-\underline{k}\downarrow} C_{\underline{k}'\uparrow} \qquad (5)$$

where $V_{ph}$ is the phonon-induced interaction, $V_{sc}$ is the screened Coulomb repulsion between conduction electrons and

$$V_l(x) = (x - x_o) \langle G^2_i / h\Omega_l \rangle \equiv (x - x_o) V_l \qquad (6)$$

Here x denotes the ratio between (H/D) and Pd (stoichiometry) and $x_o$ its value at which superconducting state first appears.

The critical transition temperature $T_c$ is computed by following the method used earlier for combined phononic and local pair (lochonic) mechanisms[12-16]. The expression for $T_c$ turns out to be

$$T_c = 1.13 \, \omega^{n_1}_{ph} \, \omega^{n_2}_l \, \text{Exp}[-1/(\lambda_{ph} + \lambda^*_l)], \qquad (7) \qquad \text{where}$$

$$n_1 = \lambda_{ph}(\lambda_{ph} + \lambda_l(x))^{-1}, \quad n_2 = \lambda_l(\lambda_{ph} + \lambda_{l(x)})^{-1} \qquad (8)$$

$$\lambda_{ph} = N(O) V_{ph}, \; \lambda^*_l = \lambda_l(x) - \mu^*, \; \lambda^{(x)}_l = N(O) V_l(x) \equiv (x - x_o)\lambda^o_l \qquad (9)$$

$\mu^*$ = the parameter arising from screened Coulomb repulsion. In Eq. (7) $\omega_{ph}$ and $\omega_l$ are the cutoff frequencies, in temperature units, for the phononic and lochonic mechanisms presented above; $N(O)$ is the density of states at the Fermi energy. For $PdH_{0.66}$, the Deby temperature has been estimated from its elastic constants to be around 285K[19]. For the present work the local optical mode of vibrations involving H/D subsystem is relevant. This is expected to be around $\omega_{ph} \simeq 300$ K. For the lochonic mechanism $\omega_l$ is estimated to lie in the range (in temperature unit) 500 K to 900 K and is connected with lochon (local pair fluctuation) modes. For other parameters we take the values

$\lambda_{ph} = 0.15$, $\lambda^0_l = 0.5$ to $0.85$, $\mu^* = 0.1$, The computed values of the critical temperature $T_c$ using some selected parameters noted above are plotted in the Figures 1 and 2 as a function of the ratio $x = H(D)/Pd$ from 0.6 to 1.6. It is found that pure phonon mechanism will give $T_c$ only a few degrees Kelvin. The combined phononic and lochonic mechanism can give much higher values particularly for $x > 1$ reaching room temperature at $x = 1.6$.

**Concluding Remarks**

A few remarks on the possible appearance of H⁻ or D⁻ (even transient) is in order at this points. It is known that the displacement of surrounding atoms can considerably reduce on site Coulomb correlation (U) and lead to onsite negative singlet pairs. Further increased loading by H(D) produces Pd vacancies in the lattice. This enhances the possibility of negative H⁻(D⁻) formation (when x is increased) in the intermediate interaction process with conduction electrons (cf..relation (A))[21].

($\underline{k}\uparrow$) – H⁺-(-$\underline{k}\downarrow$) → (o) – H⁻($\uparrow\downarrow$) – (o) → ($\underline{k'}\uparrow$) – H⁺-(-$\underline{k}'\downarrow$). (Even on Sun's photosphere the apparent opacity is due to the presence of negative hydrogen ions).

In the foregoing, a model has been suggested in which a combined mechanism involving phonons and local charged bosons (lochons) can lead to superconducting pairing of electrons in Palladium Hydride (Deuteride) systems at high enough temperatures. Phonon induced processes cannot give such high $T_c$ values. The fact that on increased loading with (H/d) there is indication of enhanced $T_c$ strongly suggests the role of hydrogen ion (H⁻/D⁻) which arise due to the conduction electron pairs existing transiently on (H⁺/D⁺) centres. It is desirable that systematic experimental work should be carried out to confirm the appearance of high $T_c$ superconductivity in Pd(H/D)$_x$ system.

**Acknowledgements**

The author should like to thank Dr. Michael McKubre of SRI International Stanford Research Institute, USA in drawing his attention to the experimental results discussed here.

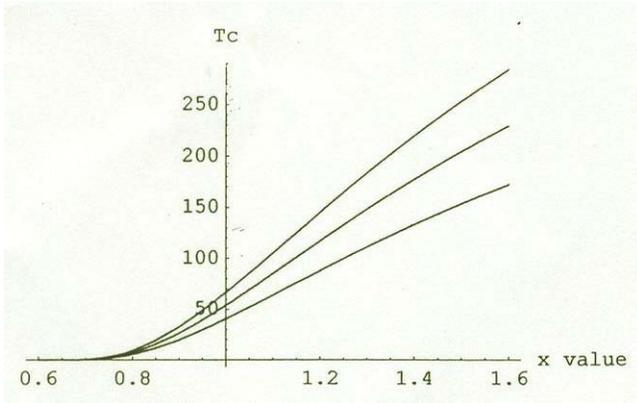

Fig. 1: $T_c$(in K) against x for three values of $\omega_l$

{500K, 700K, 900K}, $\omega_{ph} = 300K$., $\lambda_{ph} = 0.15$, $\lambda^o_l = 0.85$.

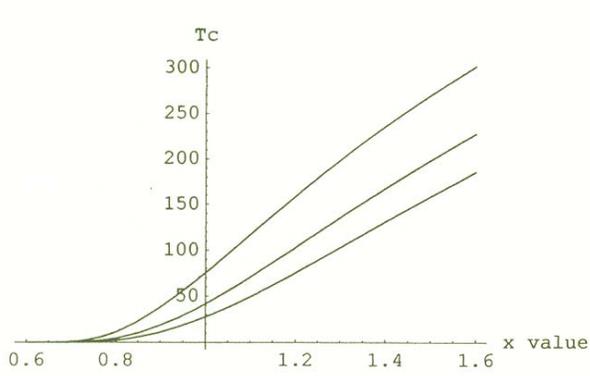

Fig. 2: $T_c$(in K) against x, For $\omega_{ph} = 300$ K, $\omega_l = 900$ K,

$\lambda_{ph} = 0.15$, $\lambda^o_l = \{0.6, 0.7, 0.9\}$